\documentclass[12pt,a4paper]{article}
\usepackage{}
\usepackage{amsfonts}
\usepackage{amssymb}
\usepackage{mathrsfs}
\usepackage{amsmath}
\usepackage{amsthm}
\usepackage{enumerate}
\usepackage{cite}
\usepackage[all]{xy}
\allowdisplaybreaks
\newtheorem{defn}{Definition}[section]
\newtheorem{thm}[defn]{Theorem}
\newtheorem{lem}[defn]{Lemma}
\newtheorem{prop}[defn]{Proposition}
\newtheorem{cor}[defn]{Corollary}
\newtheorem{eg}[defn]{Example}
\newtheorem{re}[defn]{Remark}

\newcommand\relphantom[1]{\mathrel{\phantom{#1}}}

\newcommand{\bdefn}{\begin{defn}}
\newcommand{\edefn}{\end{defn}}
\newcommand{\bthm}{\begin{thm}}
\newcommand{\ethm}{\end{thm}}
\newcommand{\blem}{\begin{lem}}
\newcommand{\elem}{\end{lem}}
\newcommand{\bprop}{\begin{prop}}
\newcommand{\eprop}{\end{prop}}
\newcommand{\bcor}{\begin{cor}}
\newcommand{\ecor}{\end{cor}}
\newcommand{\beg}{\begin{eg}}
\newcommand{\eeg}{\end{eg}}
\newcommand{\bre}{\begin{re}}
\newcommand{\ere}{\end{re}}
\newcommand{\bpf}{\begin{proof}}
\newcommand{\epf}{\end{proof}}

\newcommand{\id}{{\rm id}}

\newcommand{\Eq}{{\rm Eq}}
\newcommand{\Ker}{{\rm Ker}}
\newcommand{\K}{\mathbb{K}}

\newcommand{\benu}{\begin{enumerate}}
\newcommand{\eenu}{\end{enumerate}}
\newcommand{\bc}{\begin{center}}
\newcommand{\ec}{\end{center}}
\newcommand{\bea}{\begin{eqnarray}}
\newcommand{\eea}{\end{eqnarray}}
\newcommand{\Bea}{\begin{eqnarray*}}
\newcommand{\Eea}{\end{eqnarray*}}
\newcommand{\beq}{\begin{equation}}
\newcommand{\eeq}{\end{equation}}
\newcommand{\Beq}{\begin{equation*}}
\newcommand{\Eeq}{\end{equation*}}
\newcommand{\bspl}{\begin{split}}
\newcommand{\espl}{\end{split}}

\newcommand{\supercite}[1]{\textsuperscript{\cite{#1}}}

\begin{document}
\title{{\bf Rota-Baxter  multiplicative 3-ary Hom-Nambu-Lie algebras}}
\author{\normalsize \bf Bing Sun,  Liangyun Chen}
\date{{{\small{ School of Mathematics and Statistics,  Northeast Normal University,\\ Changchun 130024, China
 }}}} \maketitle
\date{}

\begin{abstract}

In this paper, we introduce the concepts of Rota-Baxter operators and differential operators with weights on a multiplicative $n$-ary Hom-algebra. We then focus on Rota-Baxter
multiplicative 3-ary Hom-Nambu-Lie algebras and show that they can be derived from Rota-Baxter Hom-Lie algebras, Hom-preLie algebras and Rota-Baxter commutative Hom-associative
algebras. We also explore the connections between these Rota-Baxter multiplicative 3-ary Hom-Nambu-Lie algebras.

\bigskip

\noindent {\em Key words:}  multiplicative 3-ary Hom-Nambu-Lie algebras, Rota-Baxter algebras, Hom-preLie algebras.\\
\noindent {\em Mathematics Subject Classification(2010): 16W20, 05B07}.
\end{abstract}
\renewcommand{\thefootnote}{\fnsymbol{footnote}}
\footnote[0]{ Corresponding author(L. Chen): chenly640@nenu.edu.cn.}
\footnote[0]{Supported by  NNSF of China (Nos.11171055 and
11471090),  NSF of Jilin province (No. 201115006).}
\section{Introduction}

Hom-type generalizations of $n$-ary Nambu-Lie algebras, called $n$-ary Hom-Nambu-Lie algebras, were introduced by Ataguema, Makhlouf, and Silvestrov in\cite{AJ}. Each $n$-ary Hom-Nambu-Lie
algebra has $n-1$ linear twisting maps, which appear in a twisted generalization of the $n$-ary Nambu
identity called the $n$-ary Hom-Nambu identity. If the twisting maps are all equal
to the identity, one recovers an $n$-ary Nambu-Lie algebra. The twisting maps provide a substantial
amount of freedom in manipulating Nambu-Lie algebras. For example, in\cite{AJ} it is demonstrated
that some ternary Nambu-Lie algebras can be regarded as ternary Hom-Nambu-Lie algebras with
non-identity twisting maps.

In recent years, Rota-Baxter (associative) algebras, originated from the work of G. Baxter\supercite{13} in probability and populated by the work of Cartier and Rota\supercite{14,38,39}, have also
been studied in connection with many areas of mathematics and physics, including combinatorics,
number theory, operators and quantum field theory\supercite{1,7,20,21,22,24,25,38,39}.
In particular Rota-Baxter algebras have played an important role in the Hopf algebra approach
of renormalization of perturbative quantum field theory of Connes and Kreimer\supercite{15,17,18},
as well as in the application of the renormalization method in solving divergent problems in
number theory\supercite{25,32}. Furthermore, Rota-Baxter operators on a Lie algebra are an operator
form of the classical Yang-Baxter equations and contribute to the study of integrable systems\supercite{7,8,40}. Further Rota-Baxter 3-Lie algebras are closely related to preLie algebras\supercite{BR}.

Thus it is time to study multiplicative $n$-ary Hom-Nambu-Lie algebras and Rota-Baxter algebras together to get a suitable definition
of Rota-Baxter multiplicative $n$-ary Hom-Nambu-Lie algebras. In this paper we establish
a close relationship of our definition of Rota-Baxter multiplicative 3-ary Hom-Nambu-Lie algebras with well-known concepts of
Rota-Baxter commutative Hom-associative or Hom-Lie algebras. This on one hand justifies the definition
of Rota-Baxter multiplicative 3-ary Hom-Nambu-Lie algebras and on the other hand provides a rich source of examples for
Rota-Baxter multiplicative 3-ary Hom-Nambu-Lie algebras. The concepts of differential operators and Rota-Baxter operators
with weights for general (non-associative) Hom-algebras are introduced in Section 2. The duality of
the two concepts are established. In Section 3 we extend the connections\supercite{2,9,31} from Hom-Lie
algebras and Hom-preLie algebras to multiplicative 3-ary Hom-Nambu-Lie algebras to the context of Rota-Baxter multiplicative 3-ary Hom-Nambu-Lie algebras.
In Section 4, we construct Rota-Baxter multiplicative 3-ary Hom-Nambu-Lie algebras from Rota-Baxter  commutative Hom-associative
algebras. In Section 5 we construct
Rota-Baxter multiplicative 3-ary Hom-Nambu-Lie algebras from Rota-Baxter multiplicative 3-ary Hom-Nambu-Lie algebras. We also consider the refined case of multiplicative Hom-Lie triple systems.

\section{Differential multiplicative $n$-ary Hom-algebras and Rota-Baxter multiplicative $n$-ary Hom-algebras }
We first recall some concepts related to Hom-algebras.
\bdefn\supercite{MAD}
A Hom-module is a pair $(L,\alpha)$ consisting of a $\K$-module $L$ and a linear selfmap $\alpha:L\rightarrow L$, called the twisting map.

A Hom-algebra is a triple $(L,\mu,\alpha)$ consisting of a Hom-module $(L,\alpha)$ and a bilinear map $\mu:L\otimes L\rightarrow L$.

A Hom-algebra $(L,\mu,\alpha)$ is said to be a multiplicative if for all
$x,y\in L$ we have $\alpha(\mu(x,y))=\mu(\alpha(x),\alpha(y))$.
\edefn
\bdefn\supercite{YD}
A multiplicative $n$-ary Hom-algebra $(L,\langle,\cdot \cdot \cdot,\rangle,\alpha)$ consisting of a $\K$-module $L$,
an $n$-linear map $\langle,\cdots,\rangle:L^{\otimes n} \rightarrow L$ and a linear map $\alpha:L\rightarrow L$, called the twisting map.
\edefn
\bdefn\supercite{AJ}
A multiplicative $n$-ary Hom-Nambu-Lie algebra is a triple $(L,[\cdot,\cdots,\cdot],\alpha)$ consisting of a vector space $L$ over a field $\K$, an $n$-ary multi-linear skew-symmetric operation $[\cdot,\cdots, \cdot]:L^{\otimes n}\rightarrow L$ and a linear map $\alpha$ satisfying
\begin{align}
[\alpha(y_{2}),\cdots, \alpha(y_{n}),[x_{1},\cdots, x_{n}]]=\sum_{i=1}^{n}[\alpha(x_{1}),\cdots, [x_{i}, y_{2},\cdots, y_{n}],\cdots, \alpha(x_{n})],\label{eq:1}
\end{align}
where $\alpha$ is applied to all the $n$ components outside of the inner bracket.

In particular, when $n=3$, by $\Eq.$ (\ref{eq:1}), we have
\begin{align}
 \begin{split}
[\alpha(y_{2}), \alpha(y_{3}),[x_{1},x_{2}, x_{3}]]=&[[x_{1},y_{2}, y_{3}],\alpha(x_{2}), \alpha(x_{3})]+[\alpha(x_{1}), [x_{2},y_{2},y _{3}],\alpha(x_{3})]\\
&+[\alpha(x_{1}), \alpha(x_{2}),[x_{3},y_{2}, y_{3}]]\label{eq:33}\,\,for\,\, all\, x_{1}, x_{2}, x_{3}, y_{2}, y_{3}\! \in\! L.
 \end{split}
\end{align}
 Under the skew-symmetric condition, the equation is equivalent to
\begin{align}
\begin{split}
[[x_{1},x_{2}, x_{3}],\alpha(y_{2}), \alpha(y_{3})]&=[[x_{1},y_{2}, y_{3}],\alpha(x_{2}), \alpha(x_{3})]+[[x_{2},y_{2},y_{3}],\alpha(x_{3}), \alpha(x_{1})]\\
&\relphantom{=}+[[x_{3},y_{2}, y_{3}],\alpha(x_{1}), \alpha(x_{2})]\label{eq:3}
\end{split}
\end{align}
and
\begin{align}
\begin{split}
[[x_{1},x_{2}, x_{3}],\alpha(y_{2}), \alpha(y_{3})]=&[[x_{1},y_{2}, y_{3}],\alpha(x_{2}), \alpha(x_{3})]+[\alpha(x_{1}), [x_{2},y_{2},y_{3}],\alpha(x_{3})]\\
&+[\alpha(x_{1}), \alpha(x_{2}),[x_{3},y_{2}, y_{3}]]\label{eq:2}.
\end{split}
\end{align}
\edefn
\bdefn\supercite{MA}
A Hom-associative algebra is a triple $(L,\cdot,\alpha)$ consisting of a vector space $L$  and a linear map $\alpha:L\rightarrow L$ satisfies
$$\alpha(x)\cdot (y\cdot z)=(x\cdot y)\cdot \alpha(z).\label{eq:4}$$
An endomorphism $\alpha$ is said to be an element of the centroid if $\alpha (x\cdot y)=\alpha (x)\cdot y=x\cdot \alpha(y)$ for all $x,y\in L$.

The centroid of $(L,\cdot,\alpha)$ is defined by
$$Cent(L)=\{\alpha \in {\rm End}(L):\alpha (x \cdot y)=\alpha (x) \cdot y=x \cdot \alpha (y), \,\,\forall x,y\in L\}.$$
\edefn
\bdefn\supercite{MA}
Let $(L,\cdot,\alpha)$ be a Hom-algebra and let $\lambda\in {\K}$. If a linear map $P:L\rightarrow L$ satisfies
\begin{align}
P(x)\cdot P(y)=P(P(x)\cdot y+x\cdot P(y)+\lambda x\cdot y) \,\,for\, all\, x,y\in L, \label{eq:5}
\end{align}
then $P$ is called a Rota-Baxter operator of weight $\lambda$ and $(L,\cdot,\alpha, P)$ is called a Rota-Baxter Hom-algebra of weight $\lambda$.
\edefn

\bdefn
Let $(L,\cdot,\alpha)$ be a Hom-algebra and let $\lambda\in {\K}$. If a linear map $D:L\rightarrow L$ satisfies
\beq
D(x\cdot y)=D(x)\cdot \alpha^{k}(y)+\alpha^{k}(x)\cdot D(y) \label{eq:6}
\eeq
and
\beq
D\circ \alpha=\alpha \circ D,\notag
\eeq
for all $x,y\in L$, where $k$ is a non-negative integer, then $D$ is called an $\alpha^{k}$-derivation on $(L,\cdot,\alpha)$.

More generally, a linear map $d:L\rightarrow L$ is called a derivation of weight $\lambda$ on $(L,\cdot,\alpha)$ if
\begin{align}
d(x\cdot y)=d(x)\cdot \alpha(y)+\alpha(x)\cdot d(y)+\lambda d(x)\cdot d(y) \label{eq:7}
\end{align}
and
\beq
d\circ \alpha=\alpha \circ d,\notag
\eeq
for all $x,y\in L$.
\edefn
We generalize the concepts of a Rota-Baxter operator and differential operator to multiplicative $n$-ary Hom-algebras.

\bdefn
Let $\lambda\in \K$ be fixed. A derivation of weight $\lambda$ on a multiplicative $n$-ary Hom-algebra $(L,<,\cdot \cdot \cdot,>,\alpha)$ is a linear map $d:L\rightarrow L$ such that
\begin{align}
d(\langle x_{1},\cdots, x_{n} \rangle)=\sum_{\emptyset \neq I \subseteq \{1,\cdots,n\}} \lambda^{|I|-1} \langle \bar{d}(x_{1}),\cdots, \bar{d}(x_{i}), \cdots, \bar{d}(x_{n})\rangle, \label{eq:8}
\end{align}
where
\begin{displaymath}
\bar{d}(x_{i}):=\bar{d}_{I}(x_{i}):= \left\{ \begin{array}{ll}
d(x_{i}) & \textrm{$i \in I,$}\\
\alpha(x_{i}) & \textrm{$i \notin I$}
\end{array} \right.
\end{displaymath}
for all $x_{1},\cdots, x_{n}\in L$. Then $(L,<,\cdots,>,\alpha)$ is called a differential multiplicative $n$-ary Hom-algebra of weight $\lambda$. In particular, a differential multiplicative 3-ary Hom-algebra of weight $\lambda$ is a multiplicative 3-ary Hom-algebra $(L,<,\!\cdots\!,>,\alpha)$ with a linear map $d: L\rightarrow L$
such that
\begin{align}
 \begin{split}
&d(\langle x_{1}, x_{2}, x_{3} \rangle)\\
=&\langle d(x_{1}),\alpha(x_{2}), \alpha(x_{3})\rangle+\langle \alpha(x_{1}),d(x_{2}), \alpha(x_{3})\rangle
+\langle \alpha(x_{1}),\alpha(x_{2}), d(x_{3})\rangle\\
&+\lambda \langle d(x_{1}),d(x_{2}), \alpha(x_{3})\rangle+\lambda \langle d(x_{1}),\alpha(x_{2}), d(x_{3})\rangle
+\lambda \langle \alpha(x_{1}),d(x_{2}), d(x_{3})\rangle\\
&+\lambda ^{2} \langle d(x_{1}),d(x_{2}), d(x_{3})\rangle.\label{eq:9}
 \end{split}
\end{align}
\edefn
\bdefn
Let $\lambda\in \K$ be fixed. A Rota-Baxter operator of weight $\lambda$ on a multiplicative $n$-ary Hom-algebra $(L,<,\cdot \cdot \cdot,>,\alpha)$ is a linear map $P:L\rightarrow L$ such that
\begin{align}
\langle P(x_{1}),\cdot \cdot \cdot, P(x_{n}) \rangle=P(\sum_{\emptyset \neq I \subseteq \{1,\cdots,n\}} \lambda^{|I|-1} \langle \hat{P}(x_{1}),\cdot \cdot \cdot, \hat{P}(x_{i}), \cdot \cdot \cdot, \hat{P}(x_{n})\rangle), \label{eq:10}
\end{align}
where
\begin{displaymath}
\hat{P}(x_{i}):=\hat{P}_{I}(x_{i}):= \left\{ \begin{array}{ll}
x_{i} & \textrm{$i \in I,$}\\
P(x_{i}) & \textrm{$i \notin I$}
\end{array} \right.
\end{displaymath}
for all $x_{1},\cdot \cdot \cdot, x_{n}\in L$. Then $(L,<,\cdot \cdot \cdot,>,\alpha,P)$ is called a Rota-Baxter multiplicative $n$-ary Hom-algebra of weight $\lambda$. In particular, a Rota-Baxter multiplicative 3-ary Hom-algebra of weight $\lambda$ is a multiplicative 3-ary Hom-algebra $(L,<,\cdot \cdot \cdot,>,\alpha)$ with a linear map $P: L\rightarrow L$
such that
\begin{align}
 \begin{split}
&\langle P(x_{1}), P(x_{2}), P(x_{3}) \rangle\\
=&P(\langle P(x_{1}),P(x_{2}), x_{3}\rangle+\langle P(x_{1}),x_{2}, P(x_{3})\rangle
+\langle x_{1},P(x_{2}), P(x_{3})\rangle\\
&+\lambda \langle P(x_{1}),x_{2}, x_{3}\rangle+\lambda \langle x_{1},P(x_{2}), x_{3}\rangle
+\lambda \langle x_{1},x_{2}, P(x_{3})\rangle\\
&+\lambda ^{2} \langle x_{1},x_{2}, x_{3}\rangle).\label{eq:11}
 \end{split}
\end{align}
\edefn
Specially, for $\alpha= \id$ Bai and Guo gave the differential and Rota-Baxter operators on $n$-algebra in \cite{BR}
\bthm\label{thm4}
Let $(L,<,\cdot \cdot \cdot,>,\alpha)$ be a multiplicative $n$-ary Hom-algebra over $\K$ and $\alpha$ be an algebraic automorphism. An invertible linear map $P: L\rightarrow L$ is a Rota-Baxter operator of
weight $\lambda$ on $(L,<,\cdot \cdot \cdot,>,\alpha)$ if and only if $\alpha P^{-1}$ is a differential operator of weight $\lambda$ on $(L,<,\cdot \cdot \cdot,>,\alpha)$.
\ethm
\bpf
If an invertible linear map $P: L\rightarrow L$ is a Rota-Baxter operator of weight $\lambda$, then for $x_{1},\cdots, x_{n}\in L$, let $y_{i}=P^{-1}(x_{i})$. Then by $\Eq.$ (\ref{eq:10}), we have
\Bea
\alpha P^{-1}(\langle x_{1},\cdot \cdot \cdot, x_{n} \rangle)&=&\alpha P^{-1}(\langle P(y_{1}),\cdots, P(y_{n}) \rangle)\\
&=&\alpha (\sum_{\emptyset \neq I \subseteq \{1,\cdots,n\}} \lambda^{|I|-1} \langle \hat{P}(y_{1}),\cdots, \hat{P}(y_{i}), \cdots, \hat{P}(y_{n})\rangle)\\
&=&\sum_{\emptyset \neq I \subseteq \{1,\cdots,n\}} \lambda^{|I|-1} \langle \alpha\hat{P}P^{-1}(x_{1}),\cdots, \alpha\hat{P}P^{-1}(x_{i}), \cdots, \alpha\hat{P}P^{-1}(x_{n})\rangle\\
&=&\sum_{\emptyset \neq I \subseteq \{1,\cdots,n\}} \lambda^{|I|-1} \langle \overline{\alpha P^{-1}}(x_{1}),\cdots, \overline{\alpha P^{-1}}(x_{i}), \cdots, \overline{\alpha P^{-1}}(x_{n})\rangle.
\Eea
Therefore $\alpha P^{-1}$ is a derivation of weight $\lambda$.

Conversely, let $\alpha P^{-1}$ be a derivation of weight $\lambda$. For $x_{1},\cdots, x_{n}\in A$, by $\Eq.$ (\ref{eq:8}), we have
\begin{align*}
&\alpha P^{-1}(\langle P(x_{1}),\cdot \cdot \cdot, P(x_{n}) \rangle)\\
=&\sum_{\emptyset \neq I \subseteq \{1,\cdots,n\}} \lambda^{|I|-1} \langle \overline{\alpha P^{-1}}P(x_{1}),\cdots, \overline{\alpha P^{-1}}P(x_{i}), \cdots, \overline{\alpha P^{-1}}P(x_{n})\rangle\\
=&\alpha(\sum_{\emptyset \neq I \subseteq \{1,\cdots,n\}} \lambda^{|I|-1} \langle \hat{P}(x_{1}),\cdots, \hat{P}(x_{i}), \cdots, \hat{P}(x_{n})\rangle).
\end{align*}
Therefore,
$$\langle P(x_{1}),\cdot \cdot \cdot, P(x_{n}) \rangle=P(\sum_{\emptyset \neq I \subseteq \{1,\cdots,n\}} \lambda^{|I|-1} \langle \hat{P}(x_{1}),\cdot \cdot \cdot, \hat{P}(x_{i}), \cdot \cdot \cdot, \hat{P}(x_{n})\rangle).$$
This proves the result.
\epf

\section{Rota-Baxter multiplicative 3-ary Hom-Nambu-Lie algebras from Rota-Baxter Hom-Lie algebras and Hom-preLie algebras}
We first recall the following result.
\blem\label{lem1}\supercite{AMS}
Let $(L,[,],\alpha)$ be a Hom-Lie algebra and $L^{\ast}$ be the dual space of $L$. Suppose that $f\in L^{\ast}$ satisfies $f([x,y])=0$ and $f(\alpha(x))f(y)=f(\alpha(y))f(x)$ for all $x,y\in L$. Then there is a multiplicative 3-ary Hom-Nambu-Lie algebra structure on $L$ given by
\begin{align}
[x,y,z]_{f}=f(x)[y,z]+f(y)[z,x]+f(z)[x,y]\,\,\,\,for\,all\, x,y,z\in L.\label{eq:12}
\end{align}
\elem

\bthm\label{thm1}
Let $(L,[,],\alpha,P)$ be a Rota-Baxter Hom-Lie algebra of weight $\lambda$. Suppose that $f\in L^{\ast}$ satisfies $f([x,y])=0$ and $f(\alpha(x))f(y)=f(\alpha(y))f(x)$  for all $x,y\in L$. Then $P$ is a Rota-Baxter operator on the multiplicative 3-ary Hom-Nambu-Lie algebra  $(L,[,,]_{f},\alpha)$ define in $\Eq.$ (\ref{eq:12}) if and only if $P$ satisfies
\begin{align*}
f(x)[P(y),P(z)]+f(y)[P(z),P(x)]+f(z)[P(x),P(y)]\in \Ker(P+\lambda \id)\,\,\,for\,all\, x,y,z\in L,
\end{align*}
where $\id:L\rightarrow L$ is the identity map.
\ethm
\bpf
By Lemma \ref{lem1}, $L$ is a multiplicative 3-ary Hom-Nambu-Lie algebra with the multiplication $[,,]_{f}$ defined in $\Eq.$ (\ref{eq:12}). Now for any $x,y,z\in L,$
\begin{align*}
&[P(x),P(y),P(z)]_{f}\\
=&f(P(x))[P(y),P(z)]+f(P(y))[P(z),P(x)]+f(P(z))[P(x),P(y)]\\
=&f(P(x))P([P(y),z]+[y,P(z)]+\lambda[y,z])
+f(P(y))P([P(z),x]+[z,P(x)]+\lambda[z,x])\\
&+f(P(z))P([P(x),y]+[x,P(y)]+\lambda[x,y])
\end{align*}
Applying $\Eq$s. (\ref{eq:5}) and (\ref{eq:12}), we obtain
\begin{align*}
&P([P(x),P(y),z]_{f}+[P(x),y,P(z)]_{f}+[x,P(y),P(z)]_{f}\\
&+\lambda ([P(x),y,z]_{f}+[x,P(y),z]_{f}+[x,y,P(z)]_{f})+\lambda ^{2} [x,y,z]_{f})\\
=&P(f(P(x))[P(y),z]+f(P(y))[z,P(x)]+f(z)[P(x),P(y)]\\
&+f(P(x))[y,P(z)]+f(y)[P(z),P(x)]+f(P(z))[P(x),y]\\
&+f(x)[P(y),P(z)]+f(P(y))[P(z),x]+f(P(z))[x,P(y)]\\
&+\lambda(f(P(x))[y,z]+f(y)[z,P(x)]+f(z)[P(x),y])\\
&+\lambda(f(x)[P(y),z]+f(P(y))[P(z),x]+f(z)[x,P(y)])\\
&+\lambda(f(x)[y,P(z)]+f(y)[P(z),x]+f(P(z))[x,y])\\
&+\lambda^{2}(f(x)[y,z]+f(y)[z,x]+f(z)[x,y]))\\
=&[P(x),P(y),P(z)]_{f}+P(f(x)[P(y),P(z)]+f(y)[P(z),P(x)]+f(z)[P(x),P(y)])\\
&+\lambda f(x)[P(y),P(z)]+\lambda f(y)[P(z),P(x)]+\lambda f(z)[P(x),P(y)])\\
=&[P(x),P(y),P(z)]_{f}+(P+\lambda \id)(f(x)[P(y),P(z)]+f(y)[P(z),P(x)]+f(z)[P(x),P(y)]).
\end{align*}
Then the theorem follows.
\epf

\bcor
Let $(\!L,[,],\!\alpha,\!P)$ be a Rota-Baxter Hom-Lie algebra of weight zero. Suppose that $f\in L^{\ast}$ satisfies $f([x,y])=0$ and $f(\alpha(x))f(y)=f(\alpha(y))f(x)$. Then $P$ is a Rota-Baxter operator on the multiplicative 3-ary Hom-Nambu-Lie algebra  $(L,[,,]_{f},\alpha)$ defined in $\Eq.$ (\ref{eq:12}) if and only if $P$ satisfies
\begin{align*}
[f(x)P(y)-f(y)P(x),z]+[f(y)P(z)-f(z)P(y),x]+[f(z)P(x)-f(x)P(z),y]\in \Ker P^{2},
\end{align*}
for all $x,y,z\in L$.

In particular, if $P^{2}=0$ and $f\in L^{\ast}$ satisfies $f([x,y])=0$ and $f(\alpha(x))f(y)=f(\alpha(y))f(x)$ for all $x,y\in L$. Then $P$ is a Rota-Baxter operator on the multiplicative 3-ary Hom-Nambu-Lie algebra  $(L,[,,]_{f},\alpha)$.
\ecor
\bpf
By the proof of Theorem \ref{thm1}, we have
\begin{align*}
&P([P(x),P(y),z]_{f}+[P(x),y,P(z)]_{f}+[x,P(y),P(z)]_{f})\\
=&[P(x),P(y),P(z)]_{f}+P(f(x)[P(y),P(z)]+f(y)[P(z),P(x)]+f(z)[P(x),P(y)])\\
=&[P(x),P(y),P(z)]_{f}+P^{2}([f(x)P(y)-f(y)P(x),z]+[f(y)P(z)-f(z)P(y),x]\\
&+[f(z)P(x)-f(x)P(z),y]),
\end{align*}
which proves the first statement of the corollary.

If $P^{2}=0$, then
\begin{align*}
&P([P(x),P(y),z]_{f}+[P(x),y,P(z)]_{f}+[x,P(y),P(z)]_{f})=[P(x),P(y),P(z)]_{f},
\end{align*}
which proves that $P$ is a Rota-Baxter operator on the multiplicative 3-ary Hom-Nambu-Lie algebra  $(L,[,,]_{f},\alpha)$.
\epf

\blem\label{lem6}\supercite{MA}
Let $(L,[ ,],P)$ be a Rota-Baxter Lie algebra and $\alpha:L\rightarrow L$ be a Lie algebra endomorphism commuting with $P$.
Then $(L,[ ,]_{\alpha},\alpha,P)$ is a Rota-Baxter Hom-Lie algebra, where $[ ,]_{\alpha}=\alpha\circ [ ,]$.
\elem

\bthm
Let $(L,[,],\alpha,P)$ be a Lie algebra of weight $\lambda$ and $\alpha P=P \alpha$. Suppose that $f\in L^{\ast}$ satisfies $f([x,y]_{\alpha})=0$ and $f(\alpha(x))f(y)=f(\alpha(y))f(x)$ for all $x,y\in L$. Then $(L,[,,]_{f},\alpha,P)$ is a multiplicative Rota-Baxter 3-ary Hom-Nambu-Lie algebra of weight $\lambda$ if and only if $P$ satisfies
\begin{align*}
f(x)[P(y),P(z)]_{\alpha}+f(y)[P(z),P(x)]_{\alpha}+f(z)[P(x),P(y)]_{\alpha}\in \Ker(P+\lambda \id),
\end{align*}
where $[,]_{\alpha}=\alpha\circ [,]$ and
\begin{align}
[x,y,z]_{f}=f(x)[y,z]_{\alpha}+f(y)[z,x]_{\alpha}+f(z)[x,y]_{\alpha}\,\,\,\,for\,all\, x,y,z\in L.\label{eq:34}
\end{align}
\ethm

\bpf
By Lemmas \ref{lem6} and \ref{lem1}, $(L,[ ,]_{\alpha},\alpha,P)$ is a Hom-Lie algebra of weight $\lambda$ with the multiplication $[,]_{\alpha}=\alpha\circ[,]$ and $L$ is a multiplicative 3-ary Hom-Nambu-Lie algebra with the multiplication $[,,]_{f}$ defined in $\Eq.$ (\ref{eq:34}). The rest of the proof is very similar to that of Theorem \ref{thm1}, so we will omit the details.
\epf
We now study the realizations of Rota-Baxter multiplicative 3-ary Hom-Nambu-Lie algebras by Rota-Baxter Hom-associative algebras and Rota-Baxter Hom-preLie algebras.
\bdefn\supercite{MA}
Hom-preLie algebra is a triple $(L,\ast,\alpha)$ consisting of a vector space $L$, a bilinear map $\ast :L\otimes L\rightarrow L$ and a homomorphism $\alpha$ satisfying
\begin{align*}
\alpha(x)\ast (y\ast z)-(x\ast y)\ast \alpha(z)=\alpha(y)\ast (x\ast z)-(y\ast x)\ast \alpha(z).
\end{align*}
\edefn
It is obvious that all Hom-associative algebras are Hom-preLie algebras.

For a Hom-preLie algebra $(L,\ast,\alpha)$, the commutator
\beq
[x,y]_{\ast}:=x\ast y-y\ast x\label{eq:16},
\eeq
defines a Hom-Lie algebra $G(L)=(L,[,]_{\ast},\alpha)$, called the sub-adjacent of the Hom-preLie algebra $(L,\ast,\alpha)$.

If a linear map $P:L\rightarrow L$ is a  Rota-Baxter operator of weight $\lambda$ on a Hom-preLie algebra $(L,\ast,\alpha)$, that is, $P$ satisfies
\begin{align*}
P(x)\ast P(y)=P(P(x)\ast y+x\ast P(y)+\lambda x\ast y) \,\,for\, all\, x,y\in L,
\end{align*}
then $P$ is a Rota-Baxter operator of weight $\lambda$ on its sub-adjacent Hom-Lie algebra $G(L)=(L,[,]_{\ast},\alpha)$.

\blem\label{lem2}
Let $(L,\ast,\alpha,P)$ be a Rota-Baxter Hom-preLie algebra of weight $\lambda$. Then $(L,\cdot,\alpha,P)$ is a Rota-Baxter Hom-Lie algebra of weight $\lambda$, where the multiplication $[,]_{\ast}$ is defined in $\Eq.$ (\ref{eq:16}).
\elem
\bpf
It is obvious that for any $x,y,z\in L,$
$$[x,y]_{\ast}=-[y,x]_{\ast}.$$
Furthermore, for any $x,y,z\in L$,
\begin{align*}
&[\alpha (x),[y,z]_{\ast}]_{\ast}+[\alpha (y),[z,x]_{\ast}]_{\ast}+[\alpha (z),[x,y]_{\ast}]_{\ast}\\
=&\alpha(x)\ast (y\ast z)-\alpha(x)\ast (z\ast y)-(y\ast z)\ast \alpha(x)\\
&+(z\ast y)\ast \alpha(x)+\alpha(y)\ast (z\ast x)-\alpha(y)\ast (x\ast z)\\
&-(z\ast x)\ast \alpha(y)+(x\ast z)\ast \alpha(y)+\alpha(z)\ast (x\ast y)\\
&-\alpha(z)\ast (y\ast x)-(x\ast y)\ast \alpha(z)+(y\ast x)\ast \alpha(z)\\
=&0.
\end{align*}
Moreover
\Bea
[P(x),P(y)]_{\ast}&=&P(x)\ast P(y)-P(y)\ast P(x)\\
&=&P(P(x)\ast y+x\ast P(y)+\lambda x\ast y-P(y)\ast x-y\ast P(x)-\lambda y\ast x)\\
&=&P([P(x),y]_{\ast}+[x,P(y)]_{\ast}+\lambda [x,y]_{\ast}).
\Eea
Hence, the conclusion holds.
\epf

\blem\label{lem7}\supercite{MA}
Let $(L,\circ,P)$ be a Rota-baxter associative algebra where $P$ is a Rota-baxter operator of weight $\lambda$. Let $\alpha \in Cent(L)$ and set for
$x,y\in L$
$$x\ast y=\alpha (x)\circ y.$$
Assume that $\alpha$ and $P$ commute. Then $(L,\ast,\alpha,P)$ is a Hom-associative Rota-baxter algebra.
\elem

\blem\label{lem3}
Let $(L,*,\alpha,P)$ be a Rota-Baxter Hom-preLie algebra of weight zero. Assume that $\alpha$ and $P$ commute.
  Then $(L,\cdot,\alpha,P)$ is a Rota-Baxter Hom-preLie algebra of weight zero, where the multiplication $\cdot$ is defined as
\beq
  x\cdot y:=P(x)*y-y*P(x) \,\,for\,  all\,  x,y\in L.
\eeq
\bpf
Let $(L,*,\alpha,P)$ be a Rota-Baxter Hom-preLie algebra of weight zero. For any $x,y,z\in L$, we have,
\begin{align*}
&\alpha(x)\cdot (y\cdot z)-(x\cdot y)\cdot \alpha(z)-(\alpha(y)\cdot (x\cdot z)-(y\cdot x)\cdot \alpha(z))\\
=&\alpha(P(x))\ast (P(y)\ast z-(z\ast P(y))-(P(y)\ast z-z\ast P(y))\ast \alpha(P(x))\\
&-P(P(x)\ast y-y\ast P(x))\ast \alpha(z)+\alpha(z)\ast P(P(x)\ast y-y\ast P(x))\\
&-\alpha(P(y))\ast (P(x)\ast z-(z\ast P(x))+(P(x)\ast z-z\ast P(x))\ast \alpha(P(y))\\
&+P(P(y)\ast\! x-x\ast P(y))\ast \alpha(z)-\alpha(z)\ast P(P(y)\ast x-x\ast P(y))\\
=&(P(y)\ast P(x))\ast \!\alpha(z)\!-\!(P(x)\ast\! P(y))\ast \!\alpha(z)\!+\!\alpha(z)\ast \!(P(x)\ast \!P(y))\!-\!\alpha(z)\ast\! (P(y)\ast \!P(x))\\
&\!+\!\alpha(P(x))\ast\! (P(y)\ast\! z)\!-\!\alpha(P(x))\ast \!(z\ast P(y))\!-\!(P(y)\ast\! z)\ast \!\alpha(P(x))\!+\!(z\ast \!P(y))\ast\! \alpha(P(x))\\
&\!+\!\alpha(P(y))\ast\! (P(x)\ast\! z)\!-\!\alpha(P(y))\ast \!(z\ast P(x))\!-\!(P(x)\ast \!z)\ast\! \alpha(P(y))\!+\!(z\ast\! P(x))\ast \!\alpha(P(y))\\
=&0.
\end{align*}
Furthermore,
\Bea
P(P(x)\cdot y+x\cdot P(y))&=&P(P^{2}(x)\ast y+P(x)\ast P(y))-P(y\ast P^{2}(x)+P(y)\ast P(x))\\
                          &=&P^{2}(x)\ast P(y)-P(y)\ast P^{2}(x)\\
                          &=&P(x)\cdot P(y).
\Eea
Hence, the conclusion holds.
\epf
\elem
\blem\label{lem4}
Let $(L,\cdot,\alpha,P)$ be a Rota-Baxter commutative Hom-associative algebra of weight $\lambda$,
  $D$ is an $\alpha^{0}$-derivation satisfying $DP=PD$.
  Then $(A,\ast,P)$ is a Rota-Baxter Hom-preLie algebra of weight $\lambda$, where
$$x\ast y=x\cdot D(y) \label{eq:18}\,\,for \,all\,x,y\in L.$$

In particular by Lemma \ref{lem2}, we get a  Rota-Baxter Hom-Lie algebra $(L,[,]_{\ast},\alpha)$, where
$$[x,y]_{\ast}=x\cdot D(y)-y\cdot D(x) \label{eq:19}\,\,for \,all\,x,y\in L.$$
\elem
\bpf
Let $(L,\cdot,\alpha,P)$ be a Rota-Baxter commutative Hom-associative algebra of weight $\lambda$ and $DP=PD$. Then for any $x,y,z\in L$, we have,
\begin{align*}
&\alpha(x)\ast (y\ast z)-(x\ast y)\ast \alpha(z)-(\alpha(y)\ast (x\ast z)-(y\ast x)\ast \alpha(z))\\
=&\alpha(x)\cdot(D(y)\cdot D(z))+\alpha(x)\cdot(y\cdot D^{2}(z))-(x\cdot D(y))\cdot \alpha(D(z))\\
&-\alpha(y)\cdot(D(x)\cdot D(z))-\alpha(y)\cdot(x\cdot D^{2}(z))+(y\cdot D(x))\cdot \alpha D(z)\\
=&\alpha(x)(y\cdot D^{2}(z))+(y\cdot x)\cdot \alpha(D^{2}(z))\\
=&\alpha(x)(y\cdot D^{2}(z))+(x\cdot y)\cdot \alpha(D^{2}(z))\\
=&0.
\end{align*}
Furthermore,
\begin{align*}
P(x)\ast P(y)
=&P(x)\cdot D(P(y))\\
=&P(P(x)\cdot D(y)+x\cdot P(D(y))+\lambda x\cdot D(y))\\
=&P(P(x)\ast y+x\ast P(y)+\lambda x\ast y).
\end{align*}
Hence, the conclusion holds.
\epf
\bthm\label{thm2}
Let $(L,\ast,\alpha,P)$ be a Rota-Baxter Hom-preLie algebra of weight $\lambda$. Suppose that $f\in L^{\ast}$ satisfies $f(x\ast y-y\ast x)=0$ and $f(\alpha(x))f(y)=f(\alpha(y))f(x)$  for all $x,y\in L$. For $x,y,z\in L,$ we define
\begin{align}
[x,y,z]_{f}=f(x)(y\ast z-z\ast y)+f(y)(z\ast x-x\ast z)+f(z)(x\ast y-y\ast x).\label{eq:20}
\end{align}
Then $P$ is a Rota-Baxter operator on the multiplicative 3-ary Hom-Nambu-Lie algebra  $(L,[,,]_{f},\alpha)$ if and only if
\begin{align}
\begin{split}
&f(x)(P(y)\ast P(z)-P(z)\ast P(y))+f(y)(P(z)\ast P(x)-P(x)\ast P(z))\\
&+f(z)(P(x)\ast P(y)-P(y)\ast P(x))\in \Ker(P+\lambda \id)\,\,for\,all\, x,y,z\in L.\label{eq:21}
\end{split}
\end{align}
\ethm
\bpf
By Lemma \ref{lem2}, $(L,[,]_{\ast},\alpha,P)$ is a Rota-Baxter Hom-Lie algebra of weight $\lambda$, where $[x,y]_{\ast}=x\ast y-y\ast x$.
By Lemma \ref{lem1} and Theorem \ref{thm1}, $P$ is a Rota-Baxter operator on the multiplicative 3-ary Hom-Nambu-Lie algebra  $(L,[,,]_{f},\alpha)$ if and only if $P$ satisfies
\begin{align*}
f(x)[P(y),P(z)]_{\ast}\!+\!f(y)[P(z),P(x)]_{\ast}\!+\!f(z)[P(x),P(y)]_{\ast}\!\in\! \Ker(P\!+\!\lambda \id)\,\,for\,all\, x,y,z\in L.
\end{align*}
Thus, the proof is completed.
\epf

\bthm\label{thm7}
Let $(L,\circ,P)$ be a Rota-Baxter associative algebra of weight $\lambda$, where $\alpha \in Cent(L)$. Suppose that $f\in L^{\ast}$ satisfies $f(\alpha(x)\circ y-\alpha(y)\circ x)=0$ and $f(\alpha(x))f(y)=f(\alpha(y))f(x)$  for all $x,y\in L$. For $x,y,z\in L,$ we define
\begin{align*}
[x,y,z]_{f}\!=\!f(x)(\alpha(y)\circ z\!-\!\alpha(z)\circ y)\!+\!f(y)(\alpha(z)\circ x-\alpha(x)\circ z)\!+\!f(z)(\alpha(x)\circ y-\alpha(y)\circ x).
\end{align*}
Then $P$ is a Rota-Baxter operator on the multiplicative 3-ary Hom-Nambu-Lie algebra  $(L,[,,]_{f},\alpha)$ if and only if
\begin{align}
\begin{split}
&f(x)(P(\alpha(y))\circ P(z)-P(z)\circ P(\alpha(y)))+f(y)(P(\alpha(z))\circ P(x)-P(x)\circ P(\alpha(z)))\\
&+f(z)(P(\alpha(x))\circ P(y)-P(y)\circ P(\alpha(x)))\in \Ker(P+\lambda \id)\,\,for\,all\, x,y,z\in L.\label{eq:21}
\end{split}
\end{align}
\ethm
\bpf
By Lemma \ref{lem7}, $(L,\ast,\alpha,P)$ is Rota-Baxter Hom-associative algebra of weight $\lambda$, where $x{\ast} y=\alpha(x)\ast y.$
By Lemma \ref{lem2}, $(L,[,]_{\ast},\alpha,P)$ is a Rota-Baxter Hom-Lie algebra of weight $\lambda$, where $[x,y]_{\ast}=x\ast y-y\ast x$.
By Lemma \ref{lem1} and Theorem \ref{thm1}, $P$ is a Rota-Baxter operator on the multiplicative 3-ary Hom-Nambu-Lie algebra  $(L,[,,]_{f},\alpha)$ if and only if $P$ satisfies
\begin{align*}
f(x)[P(y),P(z)]_{\ast}\!+\!f(y)[P(z),P(x)]_{\ast}\!+\!f(z)[P(x),P(y)]_{\ast}\!\in\! \Ker(P\!+\!\lambda \id)\,\,for\,all\, x,y,z\in L.
\end{align*}
Thus, the proof is completed.
\epf

\bthm
Let $(L,\ast,\alpha,P)$ be a Rota-Baxter Hom-preLie algebra of weight zero. Suppose that $f\in L^{\ast}$ satisfies $f(\alpha(x))f(y)=f(\alpha(y))f(x)$  and
\beq
f(P(x)\ast y-y\ast P(x))=f(P(y)\ast x-x\ast P(y))\,\,\,for\,all\,x,y\in L.\label{eq:22}
\eeq
Then $(L,[,,],\alpha)$ is a multiplicative 3-ary Hom-Nambu-Lie algebra with the multiplication
\begin{align}
\begin{split}
[x,y,z]:=&(f(x)P(y)-f(y)P(x))\ast z-z\ast (f(x)P(y)-f(y)P(x))\\
&+(f(z)P(x)-f(x)P(z))\ast y-y\ast (f(z)P(x)-f(x)P(z))\\
&+(f(y)P(z)-f(z)P(y))\ast x-x\ast (f(y)P(z)-f(z)P(y)).\label{eq:23}
\end{split}
\end{align}
Further, $P$ is a Rota-Baxter operator of weight zero on the multiplicative 3-ary Hom-Nambu-Lie algebra  $(L,[,,],\alpha)$ if and only if $P$ satisfies
\begin{align*}
&f(x)(P^{2}(y)\ast P^{2}(z)-P^{2}(z)\ast P^{2}(y))+f(y)(P^{2}(z)\ast P^{2}(x)-P^{2}(x)\ast P^{2}(z))\\
&+f(z)(P^{2}(x)\ast P^{2}(y)-P^{2}(y)\ast P^{2}(x))=0\,\,for\,all\, x,y,z\in L.
\end{align*}
\ethm
\bpf
By Lemma \ref{lem3}, $(L,\cdot,\alpha)$ is a Hom-preLie algebra with the multiplication
$$x\cdot y=P(x)\ast y-y\ast P(x)\,\,for \,all\,x,y\in L$$
and $P$ is a Rota-Baxter operator on the Hom-preLie algebra  $(L,\cdot,\alpha)$.

If $f\in L^{\ast}$ satisfying $f(x\cdot y-y\cdot x)=0$, that is, $f$ satisfies $\Eq.$ (\ref{eq:22}), then by Lemma \ref{thm2}, $(L,[,,],\alpha)$
is a multiplicative 3-ary Hom-Nambu-Lie algebra, where
\begin{align*}
[x,y,z]=&f(x)(y\cdot z-z\cdot y)+f(y)(z\cdot x-x\cdot z)+f(z)(x\cdot y-y\cdot x)\\
=&f(x)(P(y)\ast z-z\ast P(y)-P(z)\ast y+y\ast P(z))\\
&+f(y)(P(z)\ast x-x\ast P(z)-P(x)\ast z+z\ast P(x))\\
&+f(z)(P(x)\ast y-y\ast P(x)-P(y)\ast x+x\ast P(y))\\
=&(f(x)P(y)-f(y)P(x))\ast z-z\ast (f(x)P(y)-f(y)P(x))\\
&+(f(z)P(x)-f(x)P(z))\ast y-y\ast (f(z)P(x)-f(x)P(z))\\
&+(f(y)P(z)-f(z)P(y))\ast x-x\ast (f(y)P(z)-f(z)P(y)).
\end{align*}
Therefore, $\Eq.$ (\ref{eq:23}) holds.

By Theorem \ref{thm2},  $P$ is a Rota-Baxter operator on the multiplicative 3-ary Hom-Nambu-Lie algebra  $(L,[,,],\alpha)$ if and only if $P$ satisfies
\begin{align*}
0=&P(f(x)(P(y)\cdot P(z)-P(z)\cdot P(y))+f(y)(P(z)\cdot P(x)-P(x)\cdot P(z))\\
&+f(z)(P(x)\cdot P(y)-P(y)\cdot P(x)))\\
=&f(x)P(P^{2}(y)\ast P(z)-P(z)\ast P^{2}(y)-P^{2}(z)\ast P(y)+P(y)\ast P^{2}(z))\\
&+f(y)P(P^{2}(z)\ast P(x)-P(x)\ast P^{2}(z)-P^{2}(x)\ast P(z)+P(z)\ast P^{2}(x))\\
&+f(z)P(P^{2}(x)\ast P(y)-P(y)\ast P^{2}(x)-P^{2}(y)\ast P(x)+P(x)\ast P^{2}(y))\\
=&f(x)(P^{2}(y)\ast P^{2}(z)-P^{2}(z)\ast P^{2}(y))+f(y)(P^{2}(z)\ast P^{2}(x)-P^{2}(x)\ast P^{2}(z))\\
&+f(z)(P^{2}(x)\ast P^{2}(y)-P^{2}(y)\ast P^{2}(x)).
\end{align*}
This proves the statement.
\epf
\section{Rota-Baxter multiplicative 3-ary Hom-Nambu-Lie algebras from Rota-Baxter commutative\\ Hom-associtive algebras}
In this section, we construct Rota-Baxter multiplicative 3-ary Hom-Nambu-Lie algebras from commutative Hom-associtive algebras together with involutions and derivations. We first recall the definition of the involution.
\bdefn\supercite{WY}
Let $A$ be a commutative Hom-associtive algebra. If a linear map $\omega:A\rightarrow A$ satisfying for every $a,b\in A$,
$$\omega (ab)=\omega (a) \omega (b)$$
and
$$\omega^{2} (a)=a,$$
then $\omega$ is called an involution of $A$.
\edefn

Let $(L,\cdot,\alpha)$ be a commutative Hom-associtive algebra, $D$ be a $\alpha^{0}$-derivation and $f$ in $L^{*}$ satisfying
$f(D(x)\cdot y)=f(x\cdot D(y)),\,\,f(\alpha(x))f(y)=f(\alpha(y))f(x).$
Then by Lemmas \ref{lem1} and \ref{lem4}, $(L,[,,]_{f,D},\alpha)$ is a multiplicative 3-ary Hom-Nambu-Lie algebra, where
\begin{displaymath}
[x,y,z]_{f,D}:=
\left| \begin{array}{ccc}
f(x) & f(y) & f(z) \\
D(x) & D(y) & D(z)\\
x & y & z
\end{array}\right|
\end{displaymath}
for $x,y,z\in L$.

\bthm\label{thm3}
Let $(L,\cdot,\alpha)$ be a Rota-Baxter commutative Hom-associtive algebra of weight $\lambda$, $D$ be an $\alpha^{0}$-derivation satisfying $DP=PD$ and $f$ in $L^{*}$ satisfying
$f(D(x)\cdot y)=f(x\cdot D(y)),\,\,f(\alpha(x))f(y)=f(\alpha(y))f(x).$
Then $P$ is a Rota-Baxter operator of weight $\lambda$ on the multiplicative 3-ary Hom-Nambu-Lie algebra  $(L,[,,]_{f,D},\alpha)$
if and only if $P$ satisfies

\begin{displaymath}
\left| \begin{array}{ccc}
f(x) & f(y) & f(z) \\
DP(x) & DP(y) & DP(z)\\
P(x) & P(y) & P(z)
\end{array}\right|
\in \Ker(P+\lambda \id)\,\,for\,all \,x,y,z\in L.\label{eq:26}
\end{displaymath}
\ethm
\bpf
The result follows directly from Theorem \ref{thm1} and Lemma \ref{lem4}.
\epf

\blem\label{lem5}
Let $(L,\cdot,\alpha)$ be a commutative Hom-algebra. For a $3\times 3-matrix$ $M$, we use the notation
\begin{displaymath}
M:=
\left[ \begin{array}{ccc}
x_{1} & y_{1} & z_{1} \\
x_{2} & y_{2} & z_{2}\\
x_{3} & y_{3} & z_{3}
\end{array}\right]
=[\vec{x},\vec{y},\vec{z}]
\end{displaymath}
and the corresponding determinant, where $\vec{x},\vec{y}$ and $\vec{z}$ denote the column vectors. Let $P: L\rightarrow L$ be a Rota-Baxter operator of weight $\lambda$ and let $P(\vec{x}),P(\vec{y})$ and $P(\vec{z})$ denote the images of the column vectors. Then we have

$$| \begin{array}{ccc}
P(\vec{x}) & P(\vec{y}) & P(\vec{z})
\end{array}|
=P\left( \sum_{\emptyset \neq I \subseteq \{1,\cdots,n\}} \lambda^{|I|-1}
\left| \begin{array}{ccc}
\hat{P}(\vec{x}) & \hat{P}(\vec{y}) & \hat{P}(\vec{z})
\end{array}\right|\right).$$
\elem
\bpf
By the definition of determinants, we have
\begin{align*}
|\begin{array}{ccc}
P(\vec{x}) & P(\vec{y}) & P(\vec{z})\end{array}|=&\sum_{\sigma\in S_{3}}sgn(\sigma) P(x_{\sigma (1)}) P(y_{\sigma (2)}) P(z_{\sigma (3)})\\
=&\sum_{\sigma\in S_{3}}sgn(\sigma)P\left( \sum_{\emptyset \neq I \subseteq \{1,2,3\}} \lambda^{|I|-1} \hat{P}(x_{\sigma (1)}) \hat{P}(y_{\sigma (2)})\hat{P}(z_{\sigma (3)}) \right)\\
=&P\left( \sum_{\emptyset \neq I \subseteq \{1,2,3\}} \lambda^{|I|-1}
|\begin{array}{ccc}
\hat{P}(\vec{x}) & \hat{P}(\vec{y}) & \hat{P}(\vec{z})
\end{array}|\right)
\end{align*}
as needed.
\epf

\bthm
 Let $(L,\alpha)$ be a commutative multiplicative Hom-associtive algebra, $D$ be an $\alpha^{0}$-derivation of $L$ and $\omega:L\rightarrow L$ be an involution of $L$ satisfying
$$D(xy)=D(x)y+xD(y)\,\,for\,all\,x,y\in L,$$
$$\omega D+D \omega=0$$
and
$$\omega \alpha=\alpha \omega.$$
Then  $(L,[,,]_{\omega,D},\alpha)$ is a multiplicative 3-ary Hom-Nambu-Lie algebra with the multiplication $[,,]_{\omega,D}:L\otimes L\otimes L\rightarrow L,\,\,\forall\, x,y,z\in L,$
\begin{align}
[x,y,z]_{\omega,D}=&\left|\begin{array}{ccc}
\omega(x) & \omega(y) & \omega(z)\\ x & y & z\\ D(x) & D(y) & D(z)\end{array}\right|.\label{eq:28}
\end{align}
\ethm

\bpf
It is clear that $[,,]_{w,D}$ is a 3-ary linear skew-symmetric multiplication on $L$. Now we prove that $[,,]_{w,D}$ satisfies $\Eq.$ (\ref{eq:3}).
Since $L$ is commutative and $\omega$ is an involution of $L$ which satisfies $\omega D+D \omega=0$ and $\omega \alpha=\alpha \omega$, by $\Eq.$ (\ref{eq:28}), $\forall\,x,y,z,u,v\in L,$

\beq
\omega([x,y,z]_{\omega,D})=\left|\begin{array}{ccc}
\omega(x) & \omega(y) & \omega(z)\\ x & y & z\\ D \omega(x) & D \omega(y) & D \omega(z)\end{array}\right|,\notag
\eeq
\beq
D([x,y,z]_{\omega,D})=\left|\begin{array}{ccc}
D \omega(x) & D \omega(y) & D \omega(z)\\ x & y & z\\ D (x) & D (y) & D (z)\end{array}\right|\notag
+\left|\begin{array}{ccc}
\omega(x) & \omega(y) & \omega(z)\\ x & y & z\\ D^{2} (x) & D^{2} (y) & D^{2} (z)\end{array}\right|,
\eeq

\begin{align*}
[[x,y,z]_{\omega,D},\alpha(u),\alpha(v)]_{\omega,D}
=&\left|\begin{array}{ccc}
\omega([x,y,z]_{\omega,D}) &  \omega \alpha(u) &  \omega \alpha(v)\\
([x,y,z]_{\omega,D}) & \alpha(u) & \alpha(v)\\
D([x,y,z]_{\omega,D}) & D \alpha(u) & D \alpha(v)
\end{array}\right|\\
=&\left|\begin{array}{ccc}
\omega(x) & \omega(y) & \omega(z)\\ x & y & z\\ D \omega(x) & D \omega(y) & D \omega(z)\end{array}\right|
\left|\begin{array}{cc}
\alpha(u) & \alpha(v) \\ D \alpha(u) & D \alpha(v) \end{array}\right|\\
&+\left|\begin{array}{ccc}
\omega(x) & \omega(y) & \omega(z)\\ x & y & z\\ D (x) & D (y) &  D(z)\end{array}\right|
\left|\begin{array}{cc}
\omega\alpha(u) & \omega\alpha(v) \\ D \alpha(u) & D \alpha(v) \end{array}\right|\\
&+\left|\begin{array}{ccc}
D\omega(x) & D\omega(y) & D\omega(z)\\ x & y & z\\ D (x) & D (y) & D (z)\end{array}\right|
\left|\begin{array}{cc}
\omega\alpha(u) & \omega\alpha(v) \\  \alpha(u) &  \alpha(v) \end{array}\right|\\
&+\left|\begin{array}{ccc}
\omega(x) & \omega(y) & \omega(z)\\ x & y & z\\ D^{2}(x) & D^{2}(y) & D^{2}(z)\end{array}\right|
\left|\begin{array}{cc}
\omega\alpha(u) & \omega\alpha(v) \\ \alpha(u) &  \alpha(v) \end{array}\right|.
\end{align*}
Then we have
\begin{align*}
&[[x,u,v]_{\omega,D},\alpha(y),\alpha(z)]_{\omega,D}
+[[y,u,v]_{\omega,D},\alpha(z),\alpha(x)]_{\omega,D}
+[[z,u,v]_{\omega,D},\alpha(x),\alpha(y)]_{\omega,D}\\
=&\circlearrowleft_{x,y,z}D\omega(x)
\left|\begin{array}{cc}
\alpha(y) & \alpha(z) \\ D \alpha(y) & D \alpha(z) \end{array}\right|
\left|\begin{array}{cc}
\omega(u) & \omega(v) \\  u &  v \end{array}\right|\\
&+\circlearrowleft_{x,y,z}x
\left|\begin{array}{cc}
\omega\alpha(y) & \omega\alpha(z) \\ D \alpha(y) & D \alpha(z) \end{array}\right|
\left|\begin{array}{cc}
\omega(u) & \omega(v) \\  u &  v \end{array}\right|\\
&+\circlearrowleft_{x,y,z}D\omega(x)
\left|\begin{array}{cc}
\omega\alpha(y) & \omega\alpha(z) \\ \alpha(y) &  \alpha(z) \end{array}\right|
\left|\begin{array}{cc}
u & v \\  D(u) &  D(v) \end{array}\right|\\
&+\circlearrowleft_{x,y,z} D^{2}(x)
\left|\begin{array}{cc}
\omega\alpha(y) & \omega\alpha(z) \\  \alpha(y) &  \alpha(z) \end{array}\right|
\left|\begin{array}{cc}
\omega(u) & \omega(v) \\  u &  v \end{array}\right|\\
=&[[x,y,z]_{\omega,D},\alpha(u),\alpha(v)]_{\omega,D},
\end{align*}
where $\circlearrowleft_{x,y,z}$ is the circulation of $x,y,z,$ for example
\begin{align*}
&\circlearrowleft_{x,y,z}D\omega(x)
\left|\begin{array}{cc}
\alpha(y) & \alpha(z) \\ D \alpha(y) & D \alpha(z) \end{array}\right|\\
&=D\omega(x)
\left|\begin{array}{cc}
\alpha(y) & \alpha(z) \\ D \alpha(y) & D \alpha(z) \end{array}\right|
+D\omega(y)
\left|\begin{array}{cc}
\alpha(z) & \alpha(x) \\ D \alpha(z) & D \alpha(x) \end{array}\right|
+D\omega(z)
\left|\begin{array}{cc}
\alpha(x) & \alpha(y) \\ D \alpha(x) & D \alpha(y) \end{array}\right|.
\end{align*}
Therefore, $(L,[,,]_{\omega,D})$ is a multiplicative 3-ary Hom-Nambu-Lie algebra in the multiplication (\ref{eq:28}).
\epf

\bthm
Let $(L,\alpha,P)$ be a Rota-Baxter commutative Hom-associative algebra of weight $\lambda$, $D$ be an $\alpha^{0}$-derivation of $L$ and $\omega:L\rightarrow L$ be an involution of $L$ satisfying
$$\omega D+D \omega=0,\,\omega \alpha=\alpha \omega,\, P D=D P\,and \,P \omega=\omega P.$$
Then $P$ is a Rota-Baxter operator of weight $\lambda$ on the multiplicative 3-ary Hom-Nambu-Lie algebra $(L,[,,]_{\omega,D},\alpha)$, where $[,,]_{\omega,D}$
is defined by $\Eq.$ (\ref{eq:28}).
\ethm

\bpf
Let $x_{1},x_{2},x_{3}\in L$. By Lemma \ref{lem5} and $\Eq.$ (\ref{eq:28}), we have
\begin{align*}
[P(x_{1}),P(x_{2}),P(x_{3})]=&|\begin{array}{ccc}
\omega P(\vec{x}) & P(\vec{x}) & D P(\vec{x}) \end{array}|\\
=&|\begin{array}{ccc}
P \omega(\vec{x}) & P(\vec{x}) & P D(\vec{x}) \end{array}|\\
=&P\left( \sum_{\emptyset \neq I \subseteq \{1,2,3\}} \lambda^{|I|-1}
|\begin{array}{ccc}
\hat{P} \omega (\vec{x}) & \hat{P}(\vec{x}) & \hat{P} D (\vec{x})
\end{array}|\right)\\
=&P\left( \sum_{\emptyset \neq I \subseteq \{1,2,3\}} \lambda^{|I|-1}
|\begin{array}{ccc}
\omega  \hat{P} (\vec{x}) & \hat{P}(\vec{x}) &  D \hat{P}(\vec{x})
\end{array}|\right)\\
=&P\left( \sum_{\emptyset \neq I \subseteq \{1,2,3\}} \lambda^{|I|-1}
[\hat{P}(x_{1}),\hat{P}(x_{2}),\hat{P}(x_{3})]\right).
\end{align*}
This is what we need.
\epf

\section{Rota-Baxter multiplicative 3-ary Hom-Nambu-Lie algebras from Rota-Baxter multiplicative 3-ary\\ Hom-Nambu-Lie algebras}

Let $(L,[,,],\alpha,P)$ be a Rota-Baxter multiplicative 3-ary Hom-Nambu-Lie algebra of weight $\lambda$. Using the notation in $\Eq.$ (\ref{eq:10}), we define a ternary operation on $L$ by
\begin{align}
\begin{split}
&[x_{1},x_{2},x_{3}]_{P}\\
=&\sum_{\emptyset \neq I \subseteq \{1,2,3\}} \lambda^{|I|-1}[\hat{P}_{I}(x_{1}),\hat{P}_{I}(x_{2}),\hat{P}_{I}(x_{3})]\\
=&[P(x_{1}),P(x_{2}),x_{3}]+[P(x_{1}),x_{2},P(x_{3})]+[x_{1},P(x_{2}),P(x_{3})]\label{eq:29}\\
&+\lambda[P(x_{1}),x_{2},x_{3}]+\lambda[x_{1},P(x_{2}),x_{3}]+\lambda[x_{1},x_{2},P(x_{3})]+\lambda ^{2}[x_{1},x_{2},x_{3}],
\end{split}
\end{align}
for $x_{1},x_{2},x_{3}\in L$.

Then we have the following result.
\bthm\label{thm5}
Let $(L,[,,],\alpha,P)$ be a Rota-Baxter multiplicative 3-ary Hom-Nambu-Lie algebra of weight $\lambda$. Assume that $\alpha$ and $P$ commute. Then with $[,,]$ in $\Eq.$ (\ref{eq:29}), $(L,[,,]_{P},\alpha,P)$ is a Rota-Baxter multiplicative 3-ary Hom-Nambu-Lie algebra of weight $\lambda$.
\ethm
\bpf
First we prove that $(L,[,,]_{P},\alpha)$ is a multiplicative 3-ary Hom-Nambu-Lie algebra. It is clear that $[,,]_{P}$ is multi-linear and skew-symmetric.

Let $x_{1},x_{2},x_{3},x_{4},x_{5}\in L$. Denote $y_{1}=[x_{1},x_{2},x_{3}]_{P},y_{2}=\alpha(x_{4}),y_{3}=\alpha(x_{5})$. Then by $\Eq $s. (\ref{eq:2}), (\ref{eq:10}) and (\ref{eq:29}), we have
\begin{align*}
&[[x_{1},x_{2},x_{3}]_{P},\alpha(x_{4}),\alpha(x_{5})]_{P}\\
=&[y_{1},y_{2},y_{3}]_{P}\\
=&\sum_{\emptyset\neq I \subseteq \{1,2,3\}} \lambda^{|I|-1}[\hat{P}_{I}(y_{1}),\hat{P}_{I}(y_{2}),\hat{P}_{I}(y_{3})]\\
=&\sum_{\emptyset \neq I \subseteq \{1,2,3\},1\notin I} \lambda^{|I|-1}[P(y_{1}),\hat{P}_{I}(y_{2}),\hat{P}_{I}(y_{3})]+
\sum_{\emptyset \neq I \subseteq \{1,2,3\},1\in I} \lambda^{|I|-1}[y_{1},\hat{P}_{I}(y_{2}),\hat{P}_{I}(y_{3})]\\
=&\sum_{\emptyset \neq I \subseteq \{1,2,3\},1\notin I} \lambda^{|I|-1}[[P(x_{1}),P(x_{2}),P(x_{3})],\hat{P}_{I}(y_{2}),\hat{P}_{I}(y_{3})]\\
&+\sum_{\emptyset \neq I \subseteq \{1,2,3\},1\in I} \lambda^{|I|-1}\left[\sum_{\emptyset \neq J \subseteq \{1,2,3\}} \lambda^{|J|-1}[\hat{P}_{J}(x_{1}),\hat{P}_{J}(x_{2}),\hat{P}_{J}(x_{3})],\hat{P}_{I}(y_{2}),\hat{P}_{I}(y_{3})\right]\\
=&\sum_{\emptyset \neq K \subseteq \{1,\cdots,5\},K\cap \{1,2,3\}=\emptyset} \lambda^{|K|-1}[[\hat{P}_{K}(x_{1}),\hat{P}_{K}(x_{2}),\hat{P}_{K}(x_{3})],\alpha(\hat{P}_{K}(x_{4})),\alpha(\hat{P}_{K}(x_{5}))]\\
&+\sum_{\emptyset \neq K \subseteq \{1,\cdots,5\},K\cap \{1,2,3\}\neq \emptyset} \lambda^{|K|-1}[[\hat{P}_{K}(x_{1}),\hat{P}_{K}(x_{2}),\hat{P}_{K}(x_{3})],\alpha(\hat{P}_{K}(x_{4})),\alpha(\hat{P}_{K}(x_{5}))]\\
=&\sum_{\emptyset \neq K \subseteq \{1,\cdots,5\}} \lambda^{|K|-1}[[\hat{P}_{K}(x_{1}),\hat{P}_{K}(x_{2}),\hat{P}_{K}(x_{3})],\alpha(\hat{P}_{K}(x_{4})),\alpha(\hat{P}_{K}(x_{5}))].
\end{align*}

Since $(L,[,,],\alpha)$ is a multiplicative 3-ary Hom-Nambu-Lie algebra, for any given $\emptyset \neq K \subseteq \{1,\cdots,5\}$, we have
\begin{align*}
&[[\hat{P}_{K}(x_{1}),\hat{P}_{K}(x_{2}),\hat{P}_{K}(x_{3})],\hat{P}_{K}(x_{4}),\hat{P}_{K}(x_{5})]\\
=&[[\hat{P}_{K}(x_{1}),\hat{P}_{K}(x_{4}),\hat{P}_{K}(x_{5})],\hat{P}_{K}(x_{2}),\hat{P}_{K}(x_{3})]\\
&+[[\hat{P}_{K}(x_{2}),\hat{P}_{K}(x_{4}),\hat{P}_{K}(x_{5})],\hat{P}_{K}(x_{3}),\hat{P}_{K}(x_{1})]\\
&+[[\hat{P}_{K}(x_{3}),\hat{P}_{K}(x_{4}),\hat{P}_{K}(x_{5})],\hat{P}_{K}(x_{1}),\hat{P}_{K}(x_{2})].
\end{align*}
Thus from the above sum, we conclude that $(L,[,,]_{P},\alpha)$ is a multiplicative 3-ary Hom-Nambu-Lie algebra.

Further we have
\begin{align*}
[P(x_{1}),P(x_{2}),P(x_{3})]_{P}=&\sum_{\emptyset\neq I \subseteq \{1,2,3\}} \lambda^{|I|-1}[\hat{P}_{I}(P(x_{1})),\hat{P}_{I}(P(x_{2})),\hat{P}_{I}(P(x_{3}))]\\
=&\sum_{\emptyset\neq I \subseteq \{1,2,3\}} \lambda^{|I|-1}[P(\hat{P}_{I}(x_{1})),P(\hat{P}_{I}(x_{2})),P(\hat{P}_{I}(x_{3}))]\\
=&\sum_{\emptyset\neq I \subseteq \{1,2,3\}} \lambda^{|I|-1}P([\hat{P}_{I}(x_{1}),\hat{P}_{I}(x_{2}),\hat{P}_{I}(x_{3})]).
\end{align*}
This proves that $P$ is a  Rota-Baxter operator on $(L,[,,]_{P},\alpha)$.
\epf

\bthm\label{thm6}
Let $(L,[,,],\alpha,P)$ be a Rota-Baxter multiplicative 3-ary Hom-Nambu-Lie algebra of weight $\lambda$. Let $d$ be a differential operator of weight $\lambda$
on $(L,[,,],\alpha)$ satisfying $dP=Pd$ and $\alpha P=P \alpha$. Then $d$ is a derivation of weight $\lambda$ on the multiplicative 3-ary Hom-Nambu-Lie algebra
$(L,[,,]_{P},\alpha)$, where $[,,]_{P}$ is defined in $\Eq.$ (\ref{eq:29}).
\ethm
\bpf
Let $x_{1},x_{2},x_{3}\in L$. Using the nation in $\Eq$s. (\ref{eq:8}) and (\ref{eq:10}), we have
\begin{align*}
d([x_{1},x_{2},x_{3}]_{P})
=&\sum_{\emptyset \neq I \subseteq \{1,2,3\}} \lambda^{|I|-1}d[\hat{P}_{I}(x_{1}),\hat{P}_{I}(x_{2}),\hat{P}_{I}(x_{3})]\\
=&\sum_{\emptyset \neq I \subseteq \{1,2,3\}} \lambda^{|I|-1}\left(\sum_{\emptyset \neq J \subseteq \{1,2,3\}} \lambda^{|J|-1}[\bar{d_{J}}\hat{P}_{I}(x_{1}),\bar{d_{J}}\hat{P}_{I}(x_{2}),\bar{d_{J}}\hat{P}_{I}(x_{3})]\right)\\
=&\sum_{\emptyset \neq J \subseteq \{1,2,3\}} \lambda^{|J|-1}\left(\sum_{\emptyset \neq I \subseteq \{1,2,3\}} \lambda^{|I|-1}[\hat{P}_{I}\bar{d_{J}}(x_{1}),\hat{P}_{I}\bar{d_{J}}(x_{2}),\hat{P}_{I}\bar{d_{J}}(x_{3})]\right)\\
=&\sum_{\emptyset \neq J \subseteq \{1,2,3\}} \lambda^{|J|-1}[\bar{d_{J}}(x_{1}),\bar{d_{J}}(x_{2}),\bar{d_{J}}(x_{3})]_{P}.
\end{align*}
Therefore, $d$ is a derivation of weight $\lambda$ on the multiplicative 3-ary Hom-Nambu-Lie algebra $(L,[,,]_{P},\alpha)$.
\epf

\bcor
Let $(L,[,,],\alpha)$ be a multiplicative 3-ary Hom-Nambu-Lie algebra and $\alpha$ be an algebraic automorphism, $d$ be an invertible derivation of $L$ of weight $\lambda$. Assume that $\alpha$ and $P$ commute.
Then $(L,[,,]_{d^{-1}\alpha},\alpha)$ with $[,,]_{d^{-1}\alpha}$ defined in $\Eq.$(\ref{eq:29}) is a multiplicative 3-ary Hom-Nambu-Lie algebra. Further
\beq
[x,y,z]_{d^{-1}\alpha}=d([d^{-1}(x),d^{-1}(y),d^{-1}(z)])\,\,for \,all\,x,y,z\in L,\label{eq:30}
\eeq
and $d$ is a derivation of weight $\lambda$ on the multiplicative 3-ary Hom-Nambu-Lie algebra $(L,[,,]_{d^{-1}\alpha},\alpha)$.
\ecor
\bpf
By Theorem \ref{thm4}, $d^{-1}\alpha$ is a Rota-Baxter operator of weight $\lambda$ on the multiplicative 3-ary Hom-Nambu-Lie algebra $(L,[,,],\alpha)$. Then by Theorem \ref{thm5}, $d^{-1}\alpha$ is a Rota-Baxter  operator on the multiplicative 3-ary Hom-Nambu-Lie algebra $(L,[,,]_{d^{-1}\alpha},\alpha)$ equipped with the multiplication  $[,,]_{d^{-1}\alpha}$ defined in $\Eq.$(\ref{eq:29}). By $\Eq.$(\ref{eq:10}), we have
$$[x,y,z]_{d^{-1}\alpha}=\alpha^{-1}d(d^{-1}\alpha[x,y,z]_{d^{-1}\alpha})=d([d^{-1}(x),d^{-1}(y),d^{-1}(z)]),$$
as needed. The last statement follows from Theorem \ref{thm6}.
\epf
\bcor
Let $(L,[,],\alpha,P)$ be a Rota-Baxter Hom-Lie algebra of weight $\lambda$. Suppose that $f\in L^{\ast}$ satisfies
$f([x,y])=0$, $f(\alpha(x))f(y)=f(\alpha(y))f(x)$
and
$$(P+\lambda \id)(f(x)[P(y),P(z)]+f(y)[P(z),P(x)]+f(z)[P(x),P(y)])=0.$$
 Define
\begin{align}
[x,y,z]_{f,P}\notag
=&f(P(x))([P(y),z]+[y,P(z)]+\lambda[y,z])\\\notag
&+f(P(y))([P(z),x]+[z,P(x)]+\lambda[z,x])\\\notag
&+f(P(z))([P(x),y]+[x,P(y)]+\lambda[x,y])\\\label{eq:31}
&+f(x)([P(y),P(z)]+\lambda[P(y),z]+\lambda[y,P(z)]+\lambda^{2}[y,z])\\\notag
&+f(y)([P(z),P(x)]+\lambda[P(z),x]+\lambda[z,P(x)]+\lambda^{2}[z,x])\\\notag
&+f(z)([P(x),P(y)]+\lambda[P(x),y]+\lambda[x,P(y)]+\lambda^{2}[x,y]), \notag
\end{align}
for all $x,y,z\in L$. Then $(L,[,,]_{f,P},P)$ is a Rota-Baxter multiplicative 3-ary Hom-Nambu-Lie algebra of weight $\lambda$.
\ecor
\bpf
By Theorem \ref{thm1},
$$[x,y,z]_{f}=f(x)[y,z]+f(y)[z,x]+f(z)[x,y]$$
defines a multiplicative 3-ary Hom-Nambu-Lie algebra on $L$ for which $P$ is a Rota-Baxter operator of weight $\lambda$. Then by Theorem \ref{thm5}, the derived ternary multiplication $[,,]_{f,P}$ from $[,,]:=[,,]_{f}$ defined in $\Eq.$ (\ref{eq:29}) also equips $L$ with a multiplicative 3-ary Hom-Nambu-Lie algebra structure for which $P$ is a Rota-Baxter operator of weight $\lambda$. By direct checking, we see that $[,,]_{f,P}$ thus obtained agrees with the one defined in $\Eq.$ (\ref{eq:31}).
\epf

\bcor
Let $(L,[,],\alpha,P)$ be a Rota-Baxter Hom-Lie algebra of weight zero. Suppose that $f\in L^{\ast}$ satisfies
$f([x,y])=0$, $f(\alpha(x))f(y)=f(\alpha(y))f(x)$
and
$$P(f(x)[P(y),P(z)]+f(y)[P(z),P(x)]+f(z)[P(x),P(y)])=0.$$
Then $(L,[,,]_{P},P)$ is a Rota-Baxter multiplicative 3-ary Hom-Nambu-Lie algebra of weight zero, where
\begin{align*}
[x,y,z]_{P}
=&f(P(x))([P(y),z]+[y,P(z)])+f(P(y))([P(z),x]+[z,P(x)])\\
&+f(P(z))([P(x),y]+[x,P(y)])+f(x)[P(y),P(z)]\\
&+f(y)[P(z),P(x)]+f(z)[P(x),P(y)] \,\,\, for\, all\, x,y,z\in L.
\end{align*}
\ecor

\bthm
Let $(L,[,,],P)$ be a Rota-Baxter 3-Lie algebra and $\alpha:L\rightarrow L$ be a Lie algebra endomorphism commuting with $P$.
Then $(L,[,,]_{\alpha},\alpha,P)$ is a Rota-Baxter multiplicative 3-ary Hom-Nambu-Lie algebra, where $[,,]_{\alpha}=\alpha\circ [,,]$.
\ethm
\bpf
Observe that $[[x,y,z]_{\alpha},\alpha(u),\alpha(v)]_{\alpha}\!=\!\alpha[\alpha[x,y,z],\alpha(u),\alpha(v)]\!=\!\alpha^{2}[[x,\!y,\!z],\!u,\!v]$. Therefore the $Eq.$ (\ref{eq:1}) holds. The skew-symmetry is proved similarly.
Now we check that $P$ is still a Rota-Baxter operator for the multiplicative 3-ary Hom-Nambu-Lie algebra.
\begin{align*}
&[P(x),P(y),P(z)]_{\alpha}\\
=&\alpha ([P(x),P(y),P(z)])\\
=&\alpha (P([P(x),P(y),z]+[x,P(y),P(z)]+[P(x),y,P(z)]\\
&+\lambda[x,y,P(z)]+\lambda[P(x),y,z]+\lambda[x,P(y),z]+\lambda^{2}[x,y,z])).
\end{align*}
Since $\alpha$ and $P$ commute, we have
\begin{align*}
&[P(x),P(y),P(z)]_{\alpha}\\
=&P(\alpha[P(x),P(y),z]+\alpha[x,P(y),P(z)]+\alpha[P(x),y,P(z)]\\
&+\alpha(\lambda[x,y,P(z)])+\alpha(\lambda[P(x),y,z])+\alpha(\lambda[x,P(y),z])+\alpha(\lambda^{2}[x,y,z]))\\
=&P([P(x),P(y),z]_{\alpha}+[x,P(y),P(z)]_{\alpha}+[P(x),y,P(z)]_{\alpha}+\lambda[x,y,P(z)]_{\alpha}\\
&+\lambda[P(x),y,z]_{\alpha}+\lambda[x,P(y),z]_{\alpha}+\lambda^{2}[x,y,z]_{\alpha}).
\end{align*}
This completes the proof.
\epf
\bdefn\supercite{MAJ}
A multiplicative  Hom-Lie triple system  $(L,[,,],\alpha)$ consists
of a $\K$-vector space $L$, a trilinear map $[,,]:L\otimes L\otimes L\rightarrow L$, and a linear maps $\alpha :L \rightarrow L$
called twisted map, such that for all $x,y,z,u,v\in L$,\\
$(1)\,[x,y,y]=0,$\\
$(2)\,[x,y,z]+[y,x,z]+[z,x,y]=0,$\\
$(3)\,[[x,y,z],\alpha(u),\alpha(v)]=[[x,u,v],\alpha(y),\alpha(z)]+[\alpha(x),[y,u,v],\alpha(z)]+[\alpha(x),\alpha(y),[z,u,v]]$.
\edefn
\bthm
Let $(L,[,,],\alpha,P)$ be a Rota-Baxter multiplicative Hom-Lie triple system of weight $\lambda$. Define a ternary multiplication on $L$ by
$[,,]_{P}:L\otimes L\otimes L\rightarrow L$ in $\Eq.$ (\ref{eq:29}). Then $(L,[,,]_{P},\alpha,P)$ is a Rota-Baxter multiplicative Hom-Lie triple system.
\ethm
\bpf
It is clear that $[x,y,y]_{P}=0$ and
$$[x,y,z]_{P}+[y,z,x]_{P}+[z,x,y]_{P}=0\,\,\,for\,all\,x,y,z\in L.$$
Then the theorem follows from Theorem \ref{thm5}.
\epf
\vspace{0.3cm}

 \noindent{\bf Acknowledgements}\quad The authors would like to thank the referee for valuable comments and suggestions on this article.

\end{document}